# Nonlinear conformal-degree preserving Dirac equations

A. D. Alhaidari[†]

*Saudi Center for Theoretical Physics, Jeddah 21438, Saudi Arabia*

**Abstract**: Nonlinear Dirac equations in D+1 space-time are obtained by variation of the spinor action whose Lagrangian components have the same conformal degree and the coupling parameter of the self-interaction term is dimensionless. In 1+1 dimension, we show that these requirements result in the "conventional" quartic form of the nonlinear interaction and present the general equation for various coupling modes. These include, but not limited to, the Thirring and Gross-Neveu models. We obtain a numerical solution for the special case of the spin and pseudo-spin symmetric modes.

## I. INTRODUCTION

In nonrelativistic quantum mechanics, Schrödinger equation is a realization of de Broglie's hypothesis of the wave nature of particles [1]. Linearity of the equation is an inherent property that withstood the test of times despite various experimental attempts to observe its violation at low energy [2-5]. The prime interest in the nonlinearity of the Schrödinger equation is chiefly focused on applications in optics and condensed matter [6,7]. The absence of nonlinearity in fundamental quantum mechanics has been a source of wide interest and curiosity [8-12]. It was suggested that quantum nonlinearity could be tied to violation of Lorentz invariance at some fundamental level [13]. If so, then one would be motivated to search for nonlinearity at higher energies, short distances or strong coupling. In fact, many nonlinearities become prominent when driven by high energy sources. Therefore, evidence of quantum nonlinearity could be found straight-forwardly in the relativistic rather than nonrelativistic regime. In other words, studying the nonlinear Dirac equation could be more fruitful in this regards as compared to the nonlinear Schrödinger equation. A lot of work in the literature (both analytic and numerical) has been done over the years on the nonlinear Dirac equation; see [14] and references therein. Recently nonlinear Dirac equations were used as models to describe physical systems such as photonic crystals and Bose-Einstein condensates in optical lattices [15-17]. Well-posedness of the Dirac equation was investigated in [18,19]. Moreover, for the particular version of the equation, called the massive Thirring model, local well-posedness was studied in [20-23]. On the other hand, spectral properties of linearized Dirac operators were studied in [24] and more recently in [25-27].

On a fundamental level, solutions of nonlinear field equations have been used as models of extended particles [28-30]. For example, Soler proposed that the self-interacting 4-Fermi theory was an interesting model for extended fermions [30]. Stability of this model under dilatation was later investigated and the domain of stability was found [31]. Solitary waves in the 1+1 dimensional nonlinear Dirac equation have

---

[†] Present Address: 211 Keech Drive, Redwood City, CA 94065

−1−

been studied [32,33] in the case of massive Gross-Neveu [34] and massive Thirring [35] models. In those studies, solitary-wave solutions were found for both scalar-scalar and vector-vector interactions. The interaction between solitary waves was studied in detail for the scalar-scalar case by performing Lorentz boost on the static solutions and looking at the scattered waves [36].

Maintaining the conformal degree of the spinor field in $D+1$ space-time and choosing dimensionless coupling, we obtain in the following section the nonlinear self-interaction term in the spinor Lagrangian from which the Dirac equation is derived by variations. In 1+1 dimension, we show that these requirements result in the "conventional" quartic form of the nonlinear interaction and present the corresponding generalized equation for vector, scalar, and pseudo-scalar couplings. These include, among others, the massive Thirring and massive Gross-Neveu models. However, in 3+1 and in 2+1 dimensions, the nonlinear term has fractional powers. We limit our study to self-interaction with no external potentials. Such contribution could easily be added at a later stage. As an illustration, we calculate in Sec. III the solution of the equation in 1+1 dimension for the spin (pseudo-spin) symmetric case where the scalar interaction is equal to the positive (negative) of the vector.

## II. FORMULATION

In $D+1$ space-time, the field theoretical invariant action is $\int \mathscr{L}(x) d^n x$, where $\mathscr{L}(x)$ is the total Lagrangian and $n = D+1$. Scale invariance of the action dictates that the conformal degree of the Lagrangian is $-n$. It could be written as $\mathscr{L} = \mathscr{L}_0 + \mathscr{L}_I$, where $\mathscr{L}_0$ is the linear component and $\mathscr{L}_I$ is the self-interaction component of the Lagrangian. To obtain (by variation of the action) an equation of motion for the field $\chi(x)$ of the general form $\left(\hat{D}_0 + g \chi^\lambda\right)\chi = 0$, where $\hat{D}_0$ is the linear part, $\lambda$ is a non-zero real number and $g$ is the coupling parameter; then the nonlinear interaction becomes $\mathscr{L}_I \sim g \chi^{\lambda+2}$. Now, most renormalizable field theories (e.g., QED, QCD, Electroweak, etc.) have dimensionless coupling parameters (within the conventional relativistic units, $\hbar = c = 1$). Thus, the conformal degree of the interaction term, $\chi^{\lambda+2}$, is $-n$. In scalar (spin zero) field theories, $\mathscr{L}_0 \sim \phi \Box \phi$, where $\Box$ is the $n$-dimensional d'Alembertian operator, which is of degree $-2$. Thus, the conformal degree of the scalar field is $1 - \frac{n}{2}$ and $(\lambda+2)\left(1 - \frac{n}{2}\right) = -n$ giving $\lambda = 2\left(\frac{n}{2} - 1\right)^{-1}$. On the other hand, for spinor (spin one-half) fields, $\mathscr{L}_0 \sim i \bar{\psi} \partial\!\!\!/ \psi$ where $\partial\!\!\!/ = \sum_{\mu=0}^{n-1} \gamma^\mu \partial_\mu$ is the $n$-dimensional Dirac operator of degree $-1$ and $\{\gamma_\mu\}_{\mu=0}^{n-1}$ are the Dirac gamma matrices. Therefore, the conformal degree of the spinor field in $\frac{1}{2}(1-n)$ and $\left(\frac{\lambda}{2} + 1\right)(1-n) = -n$ giving $\lambda = 2(n-1)^{-1} = 2/D$.

From above, we see that the conformal degree of the scalar field in 3+1 space-time dimension is $-1$ and $\lambda = 2$. Therefore, the proper nonlinear interaction is the celebrated $|\phi|^4$ theory giving an equation of motion for $\phi$ with the nonlinear potential term $g|\phi|^2$. However, the conformal degree of the scalar field in 1+1 space-time is zero and $\lambda \to \infty$. Thus, the nonlinear interaction does not have a unique proper form. On the other hand, in 2+1 dimension, the conformal degree of the scalar field is $-\frac{1}{2}$ and $\lambda = 4$. Thus, a proper nonlinear interaction is the $|\phi|^6$ theory resulting in a nonlinear self-interaction part of the equation of motion of the form $g|\phi|^4$.

Now, as for the multi-component spinor field $\psi$ in $D$+1 space-time, the nonlinear self-interaction can be built using various types of couplings (e.g., scalar, vector, pseudo-scalar, etc.). These could take any of the following forms

Scalar: $\quad\quad\quad S = \bar{\psi}\psi$, $\quad\quad\quad\quad\quad\quad\quad\quad\quad\quad\quad\quad\quad\quad\quad\quad$ (1a)

Vector: $\quad\quad\quad V_\mu = \bar{\psi}\gamma_\mu\psi$, $\quad\quad\quad\quad\quad\quad\quad\quad\quad\quad\quad\quad\quad\quad\quad$ (1b)

Pseudo-Scalar: $\quad W = \bar{\psi}\gamma_5\psi$, $\quad\quad\quad\quad\quad\quad\quad\quad\quad\quad\quad\quad\quad\quad\quad$ (1c)

Pseudo-Vector: $\quad P_\mu = \bar{\psi}\gamma_\mu\gamma_5\psi$, $\quad\quad\quad\quad\quad\quad\quad\quad\quad\quad\quad\quad\quad\quad$ (1d)

Tensor: $\quad\quad\quad T_{\mu\nu} = \bar{\psi}\sigma_{\mu\nu}\psi$, $\quad\quad\quad\quad\quad\quad\quad\quad\quad\quad\quad\quad\quad\quad\quad$ (1e)

where $\bar{\psi} = \psi^\dagger \gamma_0$, $\gamma_5 = i\gamma_0\gamma_1...\gamma_D$, and $\sigma_{\mu\nu} = \frac{i}{4}[\gamma_\mu, \gamma_\nu]$. Now, in 3+1 space-time, the conformal degree of the spinor is $-\frac{3}{2}$ and $\lambda = 2/3$. Thus, possible examples of the nonlinear interaction term in the Lagrangian would be: $S(S^{1/3})$, $S(W^{1/3})$, $W(S^{1/3})$, $W(W^{1/3})$, $V^\mu(V_\mu)^{1/3}$, $P^\mu(V_\mu)^{1/3}$, $V^\mu(P_\mu)^{1/3}$, $P^\mu(P_\mu)^{1/3}$, and $T^{\mu\nu}(T_{\mu\nu})^{1/3}$, etc. For the massless case, the resulting nonlinear Dirac equation is invariant under the action of the full conformal group [37]. However, if the power of the second factor term is different from $1/3$, which will destroy the overall conformal degree, then the symmetry of the equation is reduced and invariance is only under the smaller Weyl group [37]. In 2+1 space-time, the conformal degree of the spinor is $-1$ and $\lambda = 1$. Thus, examples of the nonlinear interaction would be the same as in 3+1 dimension but with the power $1/3$ replaced by $1/2$. On the other hand, the conformal degree of the spinor in 1+1 space-time is $-\frac{1}{2}$ and $\lambda = 2$. Thus, the nonlinear interaction term can have any of the following forms[‡]

$$S^2 = (\bar{\psi}\psi)^2, \; W^2 = (\bar{\psi}\gamma_5\psi)^2, \; SW = \bar{\psi}\psi(\bar{\psi}\gamma_5\psi), \; V^2 = \bar{\psi}\gamma^\mu\psi(\bar{\psi}\gamma_\mu\psi). \quad (2)$$

These are in the conventional quartic structure, which produces the following nonlinear massive Dirac equation (in the units $\hbar = c = 1$)

$$\left[i\slashed{\partial} - m + g(\bar{\psi}\Gamma\psi)\Lambda\right]\psi = 0, \quad (3)$$

where $\Gamma$ and $\Lambda$ are Dirac matrices and $m$ is the rest mass of the particle.

Taking $\gamma_0 = \sigma_3 = \begin{pmatrix} 1 & 0 \\ 0 & -1 \end{pmatrix}$ and $\gamma_1 = i\sigma_1 = i\begin{pmatrix} 0 & 1 \\ 1 & 0 \end{pmatrix}$, Eq. (3) becomes the following nonlinear Dirac equation in 1+1 space-time with the most general coupling

$$i\partial_t \begin{pmatrix} \psi_+ \\ \psi_- \end{pmatrix} = \begin{pmatrix} m + \alpha_S S - \alpha_{SW} W + \alpha_V V & \partial_x + \alpha_W W + \alpha_{SW} S \\ -\partial_x + \alpha_W W + \alpha_{SW} S & -m - \alpha_S S + \alpha_{SW} W + \alpha_V V \end{pmatrix} \begin{pmatrix} \psi_+ \\ \psi_- \end{pmatrix}, \quad (4)$$

---

[‡] In 1+1 dimension, the tensor and pseudo-vector coupling modes are redundant.

where $\{\alpha_S, \alpha_V, \alpha_W, \alpha_{SW}\}$ are dimensionless coupling parameters. It is to be noted that we have eliminated the space component of the vector potential in the Dirac equation (4) by a $U(1)$ gauge transformation. The nonlinear self-interaction potential components become

$$S(t,x) = \bar{\psi}\psi = \psi_+^* \psi_+ - \psi_-^* \psi_- = |\psi_+|^2 - |\psi_-|^2, \tag{5a}$$

$$V(t,x) = \bar{\psi}\gamma_0\psi = \psi_+^* \psi_+ + \psi_-^* \psi_- = |\psi_+|^2 + |\psi_-|^2, \tag{5b}$$

$$W(t,x) = \bar{\psi}\gamma_5\psi = -\left(\psi_+^* \psi_- + \psi_+ \psi_-^*\right). \tag{5c}$$

We could have also made the problem more general by adding external potentials to the various coupling modes. However, we limit the present formulation to self-interaction leaving such generalization to future studies. Written in terms of the two components $\psi_\pm$, Eq. (4) with $\alpha_{SW} = 0$ reads as follows

$$i\partial_t \psi_+ = m\psi_+ + \partial_x \psi_- + \left(\alpha_+ |\psi_+|^2 + \alpha_- |\psi_-|^2\right)\psi_+ - \alpha_W \left(\psi_+ \psi_-^* + \psi_+^* \psi_-\right)\psi_-, \tag{6a}$$

$$i\partial_t \psi_- = -m\psi_- - \partial_x \psi_+ + \left(\alpha_- |\psi_+|^2 + \alpha_+ |\psi_-|^2\right)\psi_- - \alpha_W \left(\psi_+ \psi_-^* + \psi_+^* \psi_-\right)\psi_+, \tag{6b}$$

where $\alpha_\pm = \alpha_V \pm \alpha_S$.

To start, one may consider the pure vector coupling mode, where $\alpha_S = \alpha_W = 0$, which results in the following coupled nonlinear first-order partial differential equations

$$i\partial_t \psi_+ = m\psi_+ + \partial_x \psi_- + \alpha_V \left(|\psi_+|^2 + |\psi_-|^2\right)\psi_+, \tag{7a}$$

$$i\partial_t \psi_- = -m\psi_- - \partial_x \psi_+ + \alpha_V \left(|\psi_+|^2 + |\psi_-|^2\right)\psi_-. \tag{7b}$$

This case corresponds to the massive Thirring model [35]. On the other hand, the pure scalar coupling mode where $\alpha_V = \alpha_W = 0$ corresponds to the massive Gross-Neveu model [34], which reads

$$i\partial_t \psi_+ = m\psi_+ + \partial_x \psi_- + \alpha_S \left(|\psi_+|^2 - |\psi_-|^2\right)\psi_+, \tag{8a}$$

$$i\partial_t \psi_- = -m\psi_- - \partial_x \psi_+ + \alpha_S \left(|\psi_-|^2 - |\psi_+|^2\right)\psi_-. \tag{8b}$$

Generalized combination of these two models is given by Eq. (6a) and Eq. (6b) with $\alpha_W = 0$. An interesting special case of this situation is the spin and pseudo-spin symmetric model where $\alpha_W = 0$ and $\alpha_S = \pm\alpha_V$, respectively. These will result in the following equations

$$i\partial_t \psi_+ = \left(m + \alpha |\psi_\pm|^2\right)\psi_+ + \partial_x \psi_-, \tag{9a}$$

$$i\partial_t \psi_- = \left(-m + \alpha |\psi_\mp|^2\right)\psi_- - \partial_x \psi_+, \tag{9b}$$

where $\alpha_S = \pm\alpha_V = \pm\frac{1}{2}\alpha$. Finally, one may also consider the pseudo-scalar coupling mode where $\alpha_S = \alpha_V = 0$ giving the following equations

$$i\partial_t \psi_+ = m\psi_+ + \partial_x \psi_- - \alpha_W \left(\psi_+ \psi_-^* + \psi_+^* \psi_-\right)\psi_-, \tag{10a}$$

$$i\partial_t \psi_- = -m\psi_- - \partial_x \psi_+ - \alpha_W \left(\psi_+ \psi_-^* + \psi_+^* \psi_-\right)\psi_+. \tag{10b}$$

As an example, we consider in the following section the spin and pseudo-spin symmetric coupling given by Eqs. (9) and calculate their solutions numerically.

## III. A CASE STUDY

As an illustration, we obtain a numerical solution for the spin and pseudo-spin coupling modes in which the Dirac equation (4) reads as follows

$$i\partial_t \begin{pmatrix} \psi_+ \\ \psi_- \end{pmatrix} = \begin{pmatrix} m + \alpha|\psi_\pm|^2 & \partial_x \\ -\partial_x & -m + \alpha|\psi_\mp|^2 \end{pmatrix} \begin{pmatrix} \psi_+ \\ \psi_- \end{pmatrix}, \quad (11)$$

where $\alpha_S = \pm\alpha_V = \pm\frac{1}{2}\alpha$. To solve for the two components, $\psi_\pm(t,x)$, we choose the following initial state

$$\psi_+(0,x) = A_+/\cosh(\mu x), \quad (12a)$$

$$\psi_-(0,x) = A_-\tanh(\mu x)/\cosh(\mu x) \propto \frac{d}{dx}\psi_+(0,x), \quad (12b)$$

where $A_\pm$ and $\mu$ are real parameters. Figure 1 is a set of snap shots (time slices) from a video animation of the solution of the spin symmetric coupling [38]. Figure 2 gives the same but for the pseudo-spin symmetric mode. More plots are given in [38]; some show the behavior of the wavefunction after an extended length of time.

## ACKNOWLEDGEMENTS

The Author is grateful for the support provided by the Saudi Center for Theoretical Physics (SCTP). We also appreciate the help in computations offered by S. Al-Marzoug, especially in producing some of the material in [38].

**FIGURE CAPTION:**

**FIG. 1**: Solution of Eq. (11) with the top signs (the spin symmetric case) at times $t = 0$, 0.5, 1, 2, 3, 4, 5, and 6 (in units of the mass $m$). The real (imaginary) part of the wave function is shown as the red (blue) curve. The vertical axis, $\psi_\pm$, is in units of $\sqrt{m}$ whereas the horizontal $x$-axis is in units of $m^{-1}$. The top (bottom) two rows of the figure show the upper (lower) component of the wavefunction. The values of physical parameters are taken as $m = 1$, $\alpha = 0.5$, $\mu = 1$, and $A_\pm = \pm 1$.

**FIG. 2**: Same as Fig. 1 but for the pseudo-spin symmetric case corresponding to the bottom signs in Eq. (11).

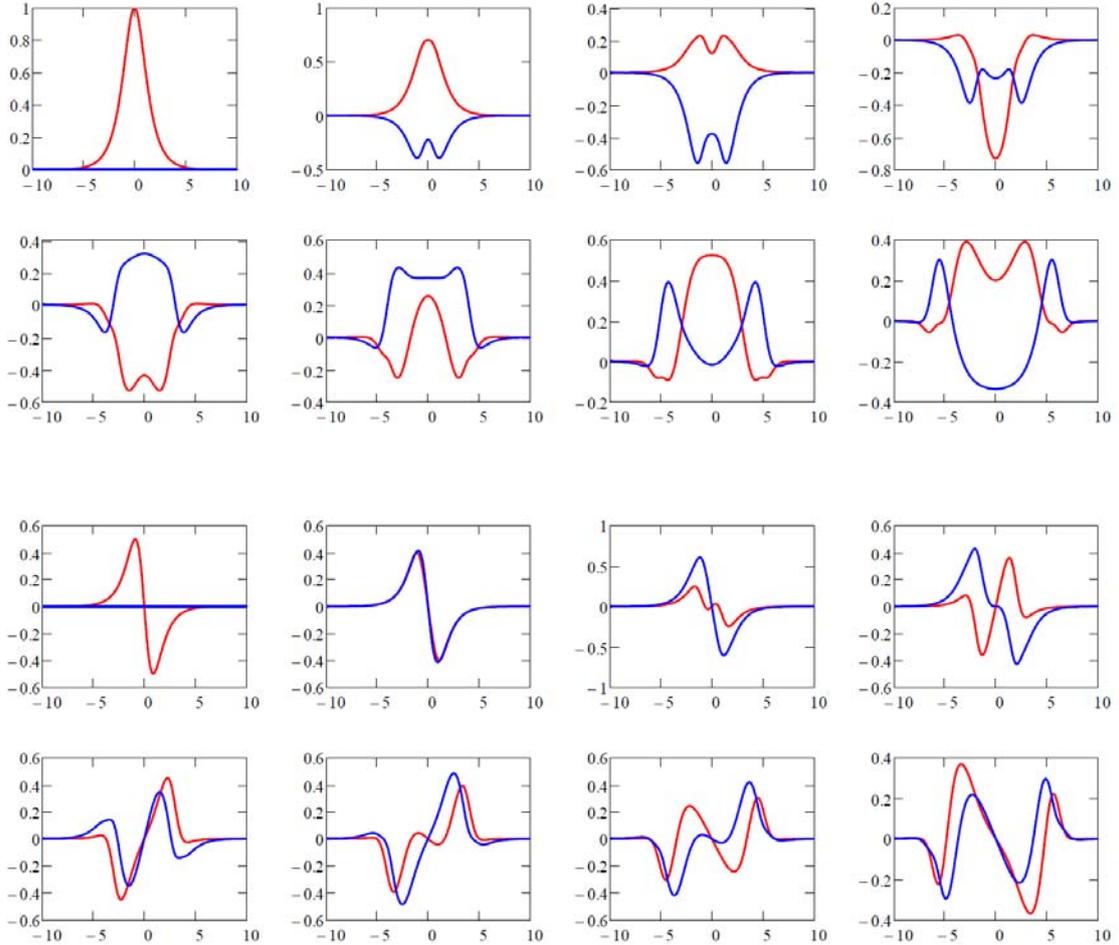

Fig. 1

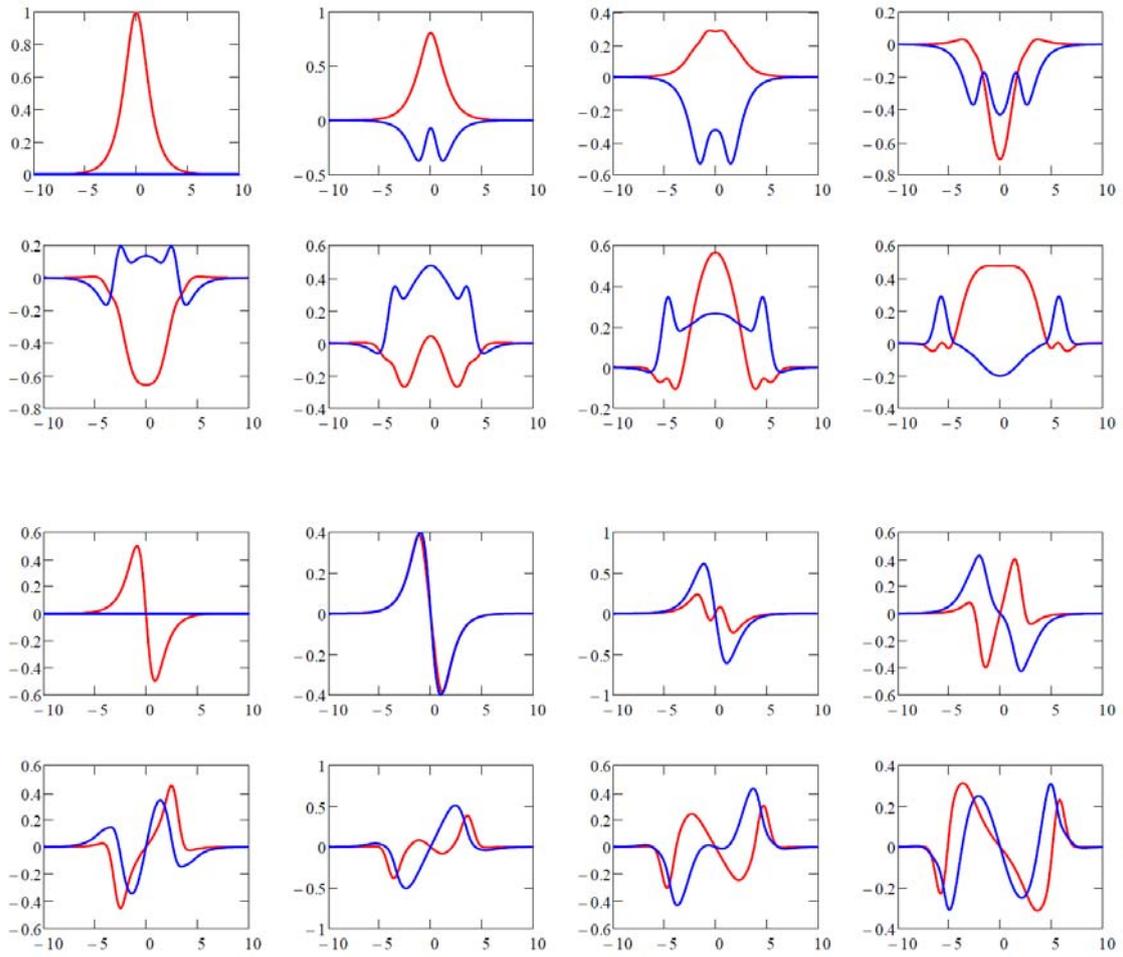

Fig. 2